\documentclass[12pt]{article}

\usepackage{float}
\usepackage{amsfonts}

\usepackage{amssymb}
\usepackage{latexsym}
\usepackage{graphicx}
\usepackage[english]{babel}
\usepackage[font={small}]{caption}

\usepackage{amsfonts}
\usepackage{latexsym}
\usepackage{graphicx}
\usepackage[english]{babel}
\usepackage{tikz}

\begin{document}
\input epsf

\def\p{\partial}
\def\h{{1\over 2}}
\def\be{\begin{equation}}
\def\bea{\begin{eqnarray}}
\def\ee{\end{equation}}
\def\eea{\end{eqnarray}}
\def\d{\partial}
\def\la{\lambda}
\def\eps{\epsilon}
\def\bb{\bigskip}
\def\mm{\medskip}
\newcommand{\dm}{\begin{displaymath}}
\newcommand{\edm}{\end{displaymath}}
\renewcommand{\b}{\tilde{B}}
\newcommand{\gm}{\Gamma}
\newcommand{\ac}[2]{\ensuremath{\{ #1, #2 \}}}
\renewcommand{\ell}{l}
\newcommand{\z}{\ell}
\newcommand{\newsection}[1]{\section{#1} \setcounter{equation}{0}}
\def\bb{$\bullet$}
\def\Qbar{{\bar Q}_1}
\def\QPbar{{\bar Q}_p}

\def\q{\quad}

\def\bn{B_\circ}

\let\a=\alpha \let\b=\beta \let\g=\gamma \let\d=\delta \let\e=\epsilon
\let\c=\chi \let\th=\theta  \let\k=\kappa
\let\l=\lambda \let\m=\mu \let\n=\nu \let\x=\xi \let\r=\rho
\let\s=\sigma \let\t=\tau
\let\vp=\varphi \let\vep=\varepsilon
\let\w=\omega      \let\G=\Gamma \let\D=\Delta \let\Th=\Theta
                     \let\P=\Pi \let\S=\Sigma

\def\h{{1\over 2}}
\def\t{\tilde}
\def\r{\rightarrow}
\def\nn{\nonumber\\}
\let\bm=\bibitem
\def\Kt{{\tilde K}}
\def\b{\bigskip}

\let\p=\partial

\begin{flushright}
\end{flushright}
\vspace{20mm}
\begin{center}
{\LARGE The nature of the gravitational vacuum \footnote{Essay awarded an honorable mention in the Gravity Research Foundation 2019 Awards for Essays on Gravitation.}
 }
\\
\vspace{18mm}
 Samir D. Mathur

\vskip .1 in

 Department of Physics\\The Ohio State University\\ Columbus,
OH 43210, USA\\mathur.16@osu.edu\\
\vspace{4mm}
\end{center}
\vspace{10mm}
\thispagestyle{empty}
\begin{abstract}

The vacuum must contain virtual fluctuations of  black hole microstates for each mass $M$. We observe that the expected suppression for $M\gg m_p$ is counteracted by the large number $Exp[S_{bek}]$ of such states. From string theory we learn that these microstates are extended objects that are resistant to compression. We argue that recognizing this `virtual extended compression-resistant' component of the gravitational vacuum is crucial for understanding gravitational physics.  Remarkably, such virtual excitations have no significant effect for observable systems like stars, but they resolve two important problems: (a) gravitational collapse is halted outside the horizon radius, removing the information paradox; (b) spacetime acquires a `stiffness' against the curving effects of vacuum energy; this ameliorates the cosmological constant problem posed by the existence of  a planck scale $\Lambda$.

\end{abstract}
\vskip 1.0 true in

\newpage
\setcounter{page}{1}


Classical general relativity is expected to fail if we encounter curvature singularities. Two singularities of special interest are the central singularity of a black hole and the big bang singularity of cosmology. Both the associated  spacetimes  exhibit horizons, and the evolution equations are similar as well: the interior of a uniform collapsing dust ball maps, under time reversal, to a region of the  dust cosmology.

But the questions we face are very different in the two cases. With black holes, we ask how information can escape from the horizon in Hawking radiation. With cosmology, we wonder why the cosmological constant is not order $\Lambda \sim l_p^{-4}$, which would curl spacetime into a planck sized ball. 

In this essay we will argue that these two very different sounding questions are resolved by a common hypothesis: {\it the vacuum of quantum gravity contains virtual black hole microstates that  are extended compression resistant objects}.

\b


{\bf {The lesson from black holes}} 

\b

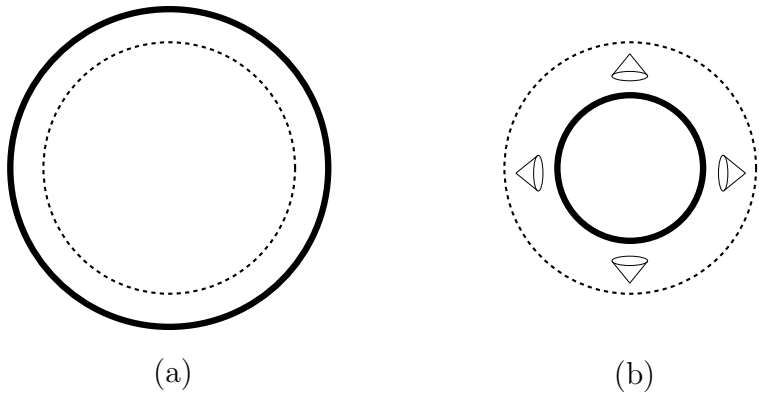
\begin{figure}
\hskip 1 in
\begin{tikzpicture}[y=0.80pt, x=.80pt, yscale=-.156000000, xscale=.156000000, inner sep=0pt, outer sep=0pt]
  \begin{scope}[shift={(0,-48.57141)}]
    \path[draw=black,dash pattern=on 1.60pt off 1.60pt,line join=miter,line
      cap=butt,miter limit=4.00,even odd rule,line width=0.800pt]
      (368.5714,580.9336) circle (10.7244cm);
    \path[draw=black,line join=miter,line cap=butt,miter limit=4.00,even odd
      rule,line width=2.400pt] (368.5714,580.9336) circle (6.2089cm);
    \path[draw=black,line join=miter,line cap=butt,miter limit=4.00,even odd
      rule,line width=0.283pt] (367.1428,302.1688) ellipse (1.5232cm and 0.3993cm);
    \path[draw=black,line join=miter,line cap=butt,miter limit=4.00,even odd
      rule,line width=0.283pt] (313.9304,300.1467) -- (367.5256,234.0937) --
      (367.5256,234.0937) -- (367.5256,234.0937) -- (367.5256,234.0937);
    \path[draw=black,line join=miter,line cap=butt,miter limit=4.00,even odd
      rule,line width=0.283pt] (421.0006,300.8207) -- (367.4054,234.7677) --
      (367.4054,234.7677) -- (367.4054,234.7677) -- (367.4054,234.7677);
    \path[xscale=1.000,yscale=-1.000,draw=black,line join=miter,line cap=butt,miter
      limit=4.00,even odd rule,line width=0.283pt] (367.1428,-862.5269) ellipse
      (1.5232cm and 0.3993cm);
    \path[draw=black,line join=miter,line cap=butt,miter limit=4.00,even odd
      rule,line width=0.283pt] (313.9304,864.5489) -- (367.5256,930.6019) --
      (367.5256,930.6019) -- (367.5256,930.6019) -- (367.5256,930.6019);
    \path[draw=black,line join=miter,line cap=butt,miter limit=4.00,even odd
      rule,line width=0.283pt] (421.0006,863.8749) -- (367.4054,929.9279) --
      (367.4054,929.9279) -- (367.4054,929.9279) -- (367.4054,929.9279);
    \path[cm={{0.0,-1.0,1.0,0.0,(0.0,0.0)}},draw=black,line join=miter,line
      cap=butt,miter limit=4.00,even odd rule,line width=0.283pt]
      (-596.6335,89.8208) ellipse (1.5232cm and 0.3993cm);
    \path[draw=black,line join=miter,line cap=butt,miter limit=4.00,even odd
      rule,line width=0.283pt] (87.7988,649.8459) -- (21.7458,596.2507) --
      (21.7458,596.2507) -- (21.7458,596.2507) -- (21.7458,596.2507);
    \path[draw=black,line join=miter,line cap=butt,miter limit=4.00,even odd
      rule,line width=0.283pt] (88.4728,542.7757) -- (22.4198,596.3709) --
      (22.4198,596.3709) -- (22.4198,596.3709) -- (22.4198,596.3709);
    \path[cm={{0.0,-1.0,-1.0,0.0,(0.0,0.0)}},draw=black,line join=miter,line
      cap=butt,miter limit=4.00,even odd rule,line width=0.283pt]
      (-596.6335,-650.1790) ellipse (1.5232cm and 0.3993cm);
    \path[draw=black,line join=miter,line cap=butt,miter limit=4.00,even odd
      rule,line width=0.283pt] (652.2010,649.8459) -- (718.2540,596.2507) --
      (718.2540,596.2507) -- (718.2540,596.2507) -- (718.2540,596.2507);
    \path[draw=black,line join=miter,line cap=butt,miter limit=4.00,even odd
      rule,line width=0.283pt] (651.5270,542.7757) -- (717.5800,596.3709) --
      (717.5800,596.3709) -- (717.5800,596.3709) -- (717.5800,596.3709);
  \end{scope}
  \begin{scope}[shift={(2.85714,-40.0)}]
    \path[draw=black,dash pattern=on 1.60pt off 1.60pt,line join=miter,line
      cap=butt,miter limit=4.00,even odd rule,line width=0.800pt]
      (-1031.4286,572.3622) circle (10.7244cm);
    \path[draw=black,line join=miter,line cap=butt,miter limit=4.00,even odd
      rule,line width=2.400pt] (-1031.4286,572.3622) circle (13.5467cm);
  \end{scope}
  \path[fill=black,line join=miter,line cap=butt,line width=0.800pt]
    (-1077.4313,1204.3622) node[above right] (text4470) {(a)};
  \path[fill=black,line join=miter,line cap=butt,line width=0.800pt]
    (319.0280,1212.9629) node[above right] (text4474) {(b)};

\end{tikzpicture}

\caption{(a) A shell of mass $M$ is collapsing towards its horizon. (b) If the shell passes through its horizon, then the information it carries is trapped inside the horizon due to the structure of light cones.} \label{f1}

\end{figure}

 Consider a collapsing shell of mass $M$. In semiclassical general relativity, this shell will pass unimpeded through its horizon radius $r_h=2GM$. The light cones turn `inwards' for $r<r_h$ (fig.\ref{f1}), so if we assume that causality holds in our theory then no effect emanating from the singularity at $r=0$ can alter the vacuum nature of the  region $0<r<r_h$. With a vacuum around the horizon, we are trapped by the information paradox \cite{hawking,cern}. 

String theory has provided  remarkable progress on this problem. In this theory we can try to explicitly construct all objects with mass $M$. For the cases studied so far, we do {\it not} find the above structure of the semiclassical hole. Instead, we find {\it fuzzballs}: horizon sized quantum objects with no horizon or singularity \cite{fuzzballs}. Fuzzballs radiate from their surface like a normal body, so there is no information puzzle.

So what alters the semiclassical collapse of the shell? In \cite{tunnel} it was argued that the collapsing shell has a probability 
\be
P\sim e^{-{4\pi ({M\over m_p})^2}}
\ee
to tunnel into a typical fuzzball microstate. While $P$ is tiny, as expected for tunneling between macroscopic objects, we must multiply by the number of fuzzball states ${\cal N}$ that we can tunnel to \cite{bek}:
\be
{\cal N}\sim e^{S_{bek}}\sim e^{{A\over 4G}}\sim e^{{4\pi ({M\over m_p})^2}}
\label{two}
\ee
We then find that the overall probability for transitioning to fuzzballs is
\be
P_{total}\sim P\times {\cal N} \sim 1
\ee
so the abnormally large value of the Bekenstein entropy creates  a violation the semiclassical approximation. 

\newpage

\b

{\bf Virtual black hole microstates}

\b

If fuzzballs exist as real objects of mass $M$, then the vacuum must contain virtual fluctuations of these objects.\footnote{The role of virtual black holes has also been considered
in other approaches; e.g. \cite{hawking1, thooft1}.}  We expect the amplitude of such a fluctuation to be tiny for $M\gg m_p$. But we argue that the large number ${\cal N}$ (eq.(\ref{two})) of  possible fuzzballs overwhelms this suppression, and makes the virtual fuzzballs an important component of the gravitational vacuum.

Two important properties of these virtual black hole microstates we obtain from their string theory constructions. First, they are extended objects, with a radius $R(M)$ close to the horizon radius $2GM$. Second, they are very compression-resistant. This can be understood from the fact that if we have energy $M$, then no more than $S_{bek}(M)$ fuzzballs should fit in a region of radius $2GM$. Compressing the fuzzballs to a smaller volume will have to raise the energy; a simple computation gives the equation of state $p=\rho$, the stiffest possible \cite{masoumi}.

Let us collect the above observations to get a picture of the vacuum of quantum gravity. This vacuum contains {\it virtual extended compression resistant objects}, which we call vecros for short. These vecros are nothing but  virtual black hole microstates, but now endowed with the properties that we have observed from explicit constructions in string theory. The largeness of their degeneracy (\ref{two}) will make them a crucial player for both the information paradox and the cosmological constant problem. 


\b

{\bf Resolving the information paradox}

\b

String theory respects causality: signals do not propagate outside the light cone. This places a stringent constraint on {\it where} a collapsing shell must transition to fuzzballs. If the shell passes unimpeded through its horizon radius $r_h$, then no effects in string theory will be able to remove the vacuum around the horizon, and will not evade the information paradox. Thus we must look for effects when the shell approaches radius $r_h$.

First consider a star of mass $M$ centered at $r=0$. The gravitational attraction of the star causes the virtual fuzzballs centered at $r=0$ to squeeze slightly; this just gives  a small change to the `vacuum polarization'. But now suppose the star collapsed to a radius $r_s<2GM$. In the region $r_s<r<2GM$ the light cones `point inwards' (fig.\ref{f1}). Any object in this region must necessarily compress to smaller radii. Virtual objects follow the same rules, so the virtual fuzzballs (vecros) would have to compress as well. But since the vecros are compression resistant, such squeezing costs a large amount of energy, which the state does not have!

What must happen instead is the following \cite{causality}. As the collapsing star reaches $r=2GM+\epsilon$, the compressive stress $F$ on virtual fuzzballs of radius $r\approx 2GM$ diverges. (This follows from the fact that we need a very large acceleration to stand just outside the radius $2GM$.) Compressing these virtual fuzzballs by a proper length $\Delta r$ transfers a large energy $E=F\Delta r$ to the virtual fuzzballs. This distorts the vacuum wavefunctional, converting the virtual fuzzballs to real on-shell fuzzball solutions. Thus semiclassical collapse halts as the star reaches its horizon radius, and we get a `string-star' instead. Since there is no horizon, we escape the information paradox.

\b

{\bf Cosmology}

\b

\begin{figure}
\begin{center}
 \includegraphics[scale=.5] {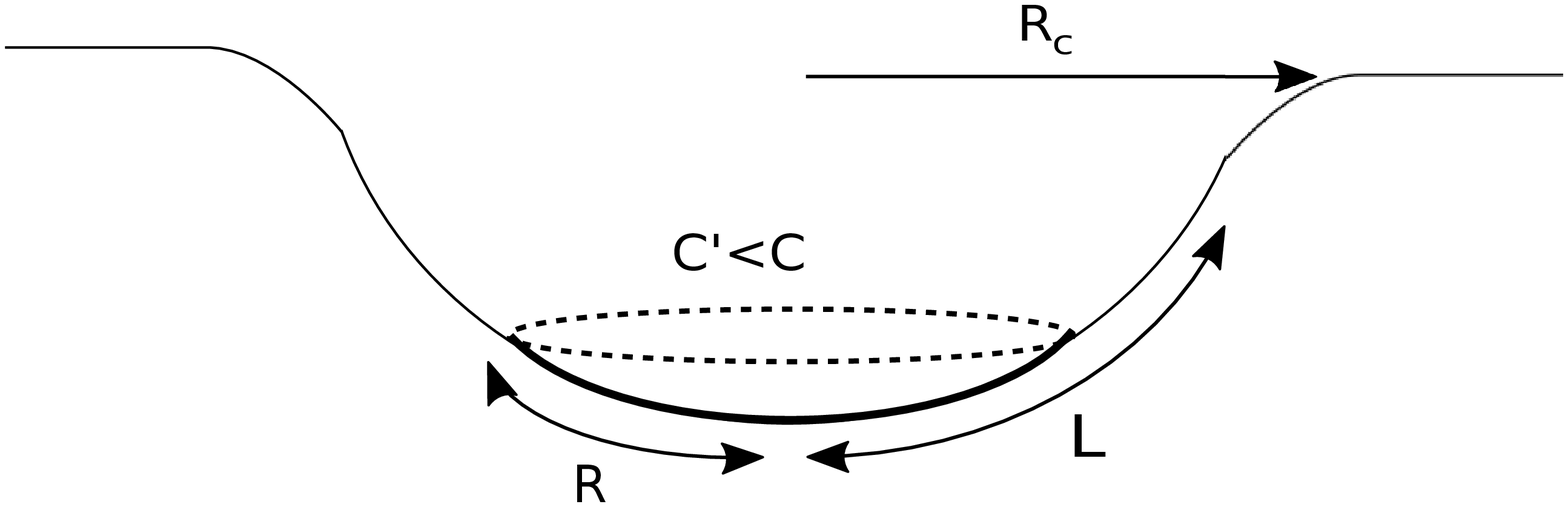}
\end{center}
\caption{A 2-d space with curvature radius $R_c$ maintained over a region of radius $L$. A disc with proper radius $R$ has a circumference $C'<C=2\pi R$; this is the compression of the vecro. } 
\label{f2}
\end{figure}

A cosmology is quite different from a black hole. Instead of a compact object we have a homogeneous infinite solution. We do not expect anything special at the cosmological horizon; rather the question is: what prevents the spacetime from curling up  everywhere in response to a vacuum energy density $\Lambda\sim l_p^{-4}$? 
We shall argue that the vecros give spacetime a stiffness which counteracts this vacuum energy.

To see this, consider 2-dimensional Euclidean space as a toy model.  Let the vacuum contain disc shaped objects (vecros) centered around each point $\vec r$  with all possible radii $0<R<\infty$. Note that when this space is flat,  a disc with radius $R$ has a circumference $C(R)=2\pi R$. 

Now suppose we bend this space to create  a hemispherical  depression centered at $r=0$ with curvature radius $R_c$ (fig.\ref{f2}). A disc centered at $r=0$ with radius $R$ now has, for the same radius $R$,  a {\it smaller} circumference $C'(R)$. The ratio $\alpha=C'/C$ is almost unity for vecros with small  $R$, bur when $R\sim R_c$, then $C'/C$ starts becoming significantly less than unity; i.e., there is significant compression of such vecros.

We model the compression resistance by taking a number $\alpha_0$ in the range $0<\alpha_0<1$ and requiring that no vecro be compressed below the factor $\alpha_0$.  From  fig.\ref{f2} we see that this puts no restriction of the curvature radius $R_c$ itself, {\it but requires that this curvature persist only for a radial distance $L$ that is $L\lesssim R_c$}. For $\alpha_0$ close to unity, we find the condition $L<L_{max}$ with
\be
L_{max}\approx R_c \left ( 6(1-\alpha_0) \right )^\h
\ee
In particular, the only space with constant curvature everywhere will be flat space.

This is the crucial point: for the actual gravitational theory, since there is no new constraint on $R_c$, there will be no change to the local gravity Lagrangian  ${\cal L}=R+a R^2+\dots$. But any curvature radius $R_c$ can only be maintained over length scales $L\lesssim R_c$; else we will encounter a stiff resistance from squeezing the virtual extended objects of radius $R\gtrsim R_c$ in the vacuum. 

A star of radius $R_{star}$ creates curvature with curvature radius $R_c\gg R_{star}$. This curvature lasts only over a radial distance $L\sim R_{star}\ll R_c$, so the vecros will feel no significant compression, and no observable effects will arise for stellar structure. The same holds for objects like galaxies or clusters where the gravitational field is weak.

 On the other hand there are two situations where we {\it do} get  $L\sim R_c$:  (i) an object compact enough to make a black hole (i.e., $R\sim GM$) and (ii) a region of a homogeneous cosmology with radius $R\sim H^{-1}$ where $H^{-1}$ is the cosmological horizon. Thus vecros will not affect the usual tests of general relativity, while they will affect both the formation of black hole horizons and dynamics at the scale of the cosmological horizon. 

More generally, we can consider a distribution function $D(R)$ giving the density of vecros with radius $R$. If $D(R)$ vanishes for $R>R_{max}$, then we can maintain a curvature radius $R_{max}$ for arbitrarily large regions, since there are no vecros with $R\gtrsim R_{max}$ to compress. This will allow spacetimes with a  nonvanishing effective cosmological constant. But we see that the value of $\Lambda$ is set by $D(R)$ rather than the vacuum energy density $\rho_0$. 

\b

{\bf Summary:}

\b

In retrospect it is not surprising that we should have to worry about virtual black hole microstates. Black holes are universal objects in all theories of gravity, and their degeneracy  $S_{bek}$  is large. String theory has told us that the microstates have radius $R\approx 2GM$ and are compression resistant;  this sets the stage for the role of these `vecros' in gravitational dynamics.

We have seen that  a shell trying to cross its horizon $r_h=2GM$ tries to crush  vecros of radius $R\approx 2GM$; this stops the collapse and generates a string star. Most solutions of the cosmological constant problem involve fine-tuning $\Lambda$. But we have argued that any uniform curvature radius $R_c$ will tend to crush vecros with $R\gtrsim R_c$; this forces flatness despite a nonvanishing vacuum energy. Thus understanding the vecro component of the gravitational vacuum may resolve many conundrums of nature.

 \section*{Acknowledgements}

I would like to thank Philip Argyres, Sumit Das, Patrick Dasgupta, Bin Guo and David Turton  for  helpful discussions.  This work is supported in part by an FQXi grant.


\newpage

\begin{thebibliography}{99}

  
 


 \bibitem{hawking}
  S.~W.~Hawking,
  Commun.\ Math.\ Phys.\  {\bf 43}, 199 (1975)
  [Erratum-ibid.\  {\bf 46}, 206 (1976)];
  S.~W.~Hawking,
  Phys.\ Rev.\  D {\bf 14}, 2460 (1976).
  
\bibitem{cern}
  S.~D.~Mathur,
  Class.\ Quant.\ Grav.\  {\bf 26}, 224001 (2009)
  [arXiv:0909.1038 [hep-th]].
  
   \bibitem{fuzzballs}
O.~Lunin and S.~D.~Mathur,
  ``AdS/CFT duality and the black hole information paradox,''
  Nucl.\ Phys.\  B {\bf 623}, 342 (2002)
  [arXiv:hep-th/0109154];
 S.~D.~Mathur,
  Fortsch.\ Phys.\  {\bf 53}, 793 (2005)
  [arXiv:hep-th/0502050];\\
I.~Kanitscheider, K.~Skenderis and M.~Taylor,
  ``Fuzzballs with internal excitations,''
  arXiv:0704.0690 [hep-th];
 I.~Bena and N.~P.~Warner,
  ``Black holes, black rings and their microstates,''
  Lect.\ Notes Phys.\  {\bf 755}, 1 (2008)
  [arXiv:hep-th/0701216];
  B.~D.~Chowdhury and A.~Virmani,
  ``Modave Lectures on Fuzzballs and Emission from the D1-D5 System,''
  arXiv:1001.1444 [hep-th].

  


   \bibitem{tunnel}
  P.~Kraus and S.~D.~Mathur,
  Int.\ J.\ Mod.\ Phys.\ D {\bf 24}, no. 12, 1543003 (2015)
  doi:10.1142/S0218271815430038
  [arXiv:1505.05078 [hep-th]];
S.~D.~Mathur,
  arXiv:0805.3716 [hep-th];
  S.~D.~Mathur,
  Int.\ J.\ Mod.\ Phys.\  D {\bf 18}, 2215 (2009)
  [arXiv:0905.4483 [hep-th]];
  I.~Bena, D.~R.~Mayerson, A.~Puhm and B.~Vercnocke,
  JHEP {\bf 1607}, 031 (2016)
  doi:10.1007/JHEP07(2016)031
  [arXiv:1512.05376 [hep-th]].
  
  \bibitem{bek}
J.~D.~Bekenstein,
Phys.\ Rev.\ D {\bf 7}, 2333 (1973).
%

\bibitem{hawking1} 
  S.~W.~Hawking,
  Phys.\ Rev.\ D {\bf 53}, 3099 (1996)
  doi:10.1103/PhysRevD.53.3099
  [hep-th/9510029].

\bibitem{thooft1} 
  G.~'t Hooft,
  Found.\ Phys.\  {\bf 48}, no. 10, 1134 (2018).
  doi:10.1007/s10701-017-0133-0

\bibitem{masoumi} 
  A.~Masoumi and S.~D.~Mathur,
  Phys.\ Rev.\ D {\bf 90}, no. 8, 084052 (2014)
  doi:10.1103/PhysRevD.90.084052
  [arXiv:1406.5798 [hep-th]];
  T.~Banks, W.~Fischler and L.~Mannelli,
  Phys.\ Rev.\ D {\bf 71}, 123514 (2005)
  doi:10.1103/PhysRevD.71.123514
  [hep-th/0408076].


\bibitem{causality} 
  S.~D.~Mathur,
  Gen.\ Rel.\ Grav.\  {\bf 51}, no. 2, 24 (2019)
  doi:10.1007/s10714-019-2505-6
  [arXiv:1703.03042 [hep-th]];
  S.~D.~Mathur,
  Int.\ J.\ Mod.\ Phys.\ D {\bf 26}, no. 12, 1742002 (2017)
  doi:10.1142/S0218271817420020
  [arXiv:1705.06407 [hep-th]];
   




   

\end{thebibliography}
\end{document}